\documentclass[aps,prb,showpacs,superscriptaddress,twocolumn]{revtex4}
\usepackage{amsmath}
\usepackage{amsfonts}
\usepackage{graphicx}

\begin{document}

\title{Quantum rings as electron spin beam splitters}
\author{P\'{e}ter F\"{o}ldi}
\affiliation{Department of Theoretical Physics, University of Szeged, Tisza
Lajos k\"{o}r\'{u}t 84, H-6720 Szeged, Hungary}
\author{Orsolya K\'{a}lm\'{a}n}
\affiliation{Department of Theoretical Physics, University of Szeged, Tisza
Lajos k\"{o}r\'{u}t 84, H-6720 Szeged, Hungary}
\author{Mih\'{a}ly G. Benedict}\email{benedict@physx.u-szeged.hu}
\affiliation{Department of Theoretical Physics, University of Szeged, Tisza
Lajos k\"{o}r\'{u}t 84, H-6720 Szeged, Hungary}
\author{F. M. Peeters}\email{francois.peeters@ua.ac.be}
\affiliation{Departement Fysica, Universiteit Antwerpen, Groenenborgerlaan
171, B-2020 Antwerpen, Belgium}

\begin{abstract}
Quantum interference and spin-orbit interaction in a one-dimensional
mesoscopic semiconductor ring with one input and two output leads can act as a
spin beam splitter. Different polarization can be achieved in the two output
channels from an originally totally unpolarized incoming spin state, very much
like in a Stern-Gerlach apparatus. We determine the relevant parameters such
that the device has unit efficiency.
\end{abstract}

\maketitle

The Stern Gerlach experiment, where spatial and spin degrees of freedom become
intertwined, has been playing a fundamental role in the conceptual foundations
of Quantum Mechanics. Still, soon after the discovery of this effect, it was
pointed out by Bohr and Mott \cite{MM49} that, in contrast to atoms, electrons
can not be spin-polarized in an inhomogeneous magnetic field. The recent
spectacular development of spin electronics (spintronics) \cite{ZFS04} in low
dimensional semiconductor structures offers a new way of manipulating spin
degrees of freedom. Quantum rings made of semiconducting material
\cite{VKDM04} exhibiting Rashba-type \cite{Ras} spin-orbit interaction (SOI)
have been shown to be especially important due to their remarkable spin
transformation properties \cite{NATE97,NMT99,MPV,FR04,FMBP05a,ZX05,KN04}.

\begin{figure}[tbh]
\begin{center}
\includegraphics*[width=7.0cm]{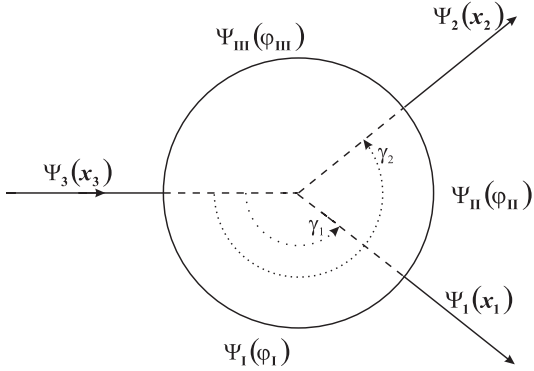}
\vspace{-0.4cm}
\end{center}
\caption{ The geometry of the device and the relevant wave functions in the
different domains. } \label{ringfig}
\end{figure}

In the present paper we propose a device that can be considered to a large
extent a {\it spintronic analogue of the Stern-Gerlach apparatus}: the
incoming electrons are forced to split into two different spatial parts by the
geometrical construction of the semiconductor device, see Fig.~\ref{ringfig}.
Due to spin-sensitive quantum interference \cite{SN05,KMGA05,CCZ04} and
spin-orbit interaction, electrons that are initially in a totally unpolarized
spin state become polarized at the outputs with different spin directions. A
similar polarizing effect has been predicted in a Y-shaped conductor as a
consequence of scattering on impurities\cite{P04} (which is a different
physical mechanism from the coherent spin transfer to be discussed here) or
because of the presence of SOI in a localized area around the
junction.\cite{KK01} There are important proposals considering four terminal
devices \cite{IDA03,G02} as well, where the strength of the SOI is assumed to
be different in the two arms of the interferometer. In our model SOI is
uniform in the ring and absent in the leads. However, the latter requirement
is not crucial, its purpose is to demonstrate clearly the role of the ring
itself, while the effects caused by SOI in the leads can be included in a
straightforward way. As our treatment is based on an exact, analytic solution
of the spin dependent transport problem, it allows us to determine for which
parameters the device is reflectionless, i.e, perfect polarization at the
outputs takes place without losses.
\bigskip

We consider a ring \cite{ALG93} of radius $a$ in the $x-y$ plane and assume a
tunable static electric field in the $z$ direction controlling the strength of
the spin-orbit interaction characterized by the parameter
$\alpha$.\cite{NATE97} The Hamiltonian \cite{Meijer,MPV} in the presence of
spin-orbit interaction for a charged particle of effective mass $m^{\ast }$ is
given by
\begin{equation}
H=\hbar \Omega \left[ \left( -i\frac{\partial }{\partial \varphi }+\frac{%
\omega }{2\Omega }(\sigma _{x}\cos \varphi +\sigma _{y}\sin \varphi )\right)
^{2}-\frac{\omega ^{2}}{4\Omega ^{2}}\right] ,  \label{Ham}
\end{equation}%
where $\varphi $ is the azimuthal angle of a point on the ring, $\hbar\Omega
=\hbar ^{2}/2m^{\ast }a^{2}$ is the dimensionless kinetic energy of the
charged particle and $\omega$ =$\alpha /\hbar a$ is the frequency associated
with the SOI. According to Ref.~[\onlinecite{FMBP05a}], in the $\left\vert
\uparrow\right\rangle$, $\left\vert \downarrow\right\rangle $ eigenbasis of
the $z$ component of the spin, the eigenstates of $H$ read:
\begin{equation}
\psi (\kappa ,\varphi )=e^{i\kappa \varphi }\binom{e^{-i\varphi /2}u(\kappa )%
}{e^{i\varphi /2}v(\kappa )}.  \label{est}
\end{equation}%
The corresponding energy eigenvalues are
\begin{equation}
E=\hbar \Omega \left[ \kappa ^{2}-\mu \kappa w+1/4\right] ,\quad \mu =\pm 1,
\label{En}
\end{equation}%
with $w=\sqrt{1+(\omega ^{2}/\Omega ^{2})}$. The spinors in (\ref{est}) are
simultaneous eigenvectors of $H$, of the $z$ component of the total angular
momentum: $K=L_{z}+$ $S_{z}$, and of the spin
operator pointing in the direction determined by the angles $\theta $ and $%
\varphi $:
\begin{equation}S_{\theta \varphi }=S_{x}\sin \theta \cos \varphi
+S_{y}\sin \theta \sin \varphi +S_{z}\cos \theta ,
\end{equation} where
$\theta $ is given by the
constant $\tan \theta =-\omega /\Omega :$%
\begin{equation}
K\psi (\kappa ,\varphi )=\kappa _{j}^{\mu }\psi (\kappa _{j}^{\mu },\varphi
),\quad _{{}}S_{\theta \varphi }\psi (\kappa _{j}^{\mu },\varphi )=\frac{\mu
}{2} \psi(\kappa _{j}^{\mu },\varphi ). \label{eig}
\end{equation}%
From geometrical point of view, the second eigenvalue equation above means
that the direction of the spinors (\ref{est}) are either parallel or
antiparallel with the conserved (position dependent) direction defined by
$S_{\theta \varphi }$. Therefore, the expectation value of the vector
$\vec{S}$ in these states rotates around the $z$ direction making always an
angle $\theta $ with it, while $\varphi $ is the actual azimuth along the
ring.

In a closed ring $\kappa \pm 1/2$ should be integer, but the presence of the
leads connected to the ring lifts this restriction: the energy is a continuous
variable, and then the possible values of $\kappa $ are the solutions of
Eq.~(\ref{En}):
\begin{equation}
\kappa _{j}^{\mu }=\mu (w/2+(-1)^{j}q),\qquad j=1,2,\quad \mu =\pm 1,
\end{equation}%
where $q=\sqrt{(\omega /2\Omega )^{2}+E/\hbar \Omega }$. The energy
eigenvalues are fourfold degenerate, they can be classified \cite{FMBP05a} by
the quantum numbers $\kappa$ and $\mu$. The ratio of the components of the
eigenvectors (\ref{est}) is determined by
$ {v(\kappa _{j}^{\mu })}/{u(\kappa _{j}^{\mu })}=(\tan {\theta /2})_{\mu }%
={\Omega }/{\omega }\left( 1-\mu w\right). $

The stationary states of the complete problem including the ring as well as
the leads, can be obtained by fitting the solutions corresponding to the
different domains. Using local coordinates as shown in Fig.~\ref{ringfig}, the
incoming wave, $\Psi _{3}(x_{3})$, and the outgoing waves $\Psi _{1}(x_{1}),$
$\Psi _{2}(x_{2})$ are built up as linear combinations of spinors with spatial
dependence $e^{ikx}$ etc. corresponding to $E={\hbar ^{2}k^{2}}/{2m^{\ast }}$:
\begin{subequations}
\begin{eqnarray}
\Psi _{3}\left( x_{3}\right) &=&
\begin{pmatrix}
f_{\uparrow} \\
f_{\downarrow}%
\end{pmatrix}
e^{ikx_{3}}+
\begin{pmatrix}
r_{\uparrow} \\
r_{\downarrow}
\end{pmatrix}
e^{-ikx_{3}}, \\  \Psi _{n}\left( x_{n}\right) &=&
\begin{pmatrix}
t_{\uparrow}^{n} \\
t_{\downarrow}^{n}
\end{pmatrix}
e^{ikx_{n}},
\end{eqnarray}
\end{subequations} where $n={1, 2}.$
The wave functions belonging to the same energy $E$ in all the three sections
of the ring can be written as linear combinations of four eigenspinors:
\begin{equation}
\Psi _{i}(\varphi_{i} )=\sum_{\substack{ j=1,2  \\ \mu =\pm 1 }}a_{i j \mu
}\psi (\kappa _{j}^{\mu },\varphi_{i} ), \label{inthering}
\end{equation}%
with $i=I,II,III$ identifying the sections. This superposition is no more an
eigenvector of $S_{\theta \varphi },$ as it contains states with both $\mu
=\pm 1,$ and their coefficients are different in general. Additionally,
spatial interference of terms describing clockwise and anticlockwise motion
plays an essential role in determining the spin direction in the ring.
Therefore the position dependencies of the corresponding spin expectation
values $\langle \Psi _{i}\vert \vec{S} \vert \Psi _{i}\rangle$ in the arms are
more complicated than the simple precession of the eigenvectors given in
Eq.~(\ref{est}). Nevertheless, this change of the spin direction along the
ring can be calculated without difficulty, if one determines the values of the
coefficients $a_{ij\mu },$ which can be done using the boundary conditions, to
be discussed now.

Figure \ref{ringfig} indicates the wave functions to be fitted at different
junctions: e.g.~the incoming wave at $x_{3}=0$ should be fitted to $\Psi _{I}$
at $\varphi_{I} =0 $ and to $\Psi _{III}$ at $\varphi _{III}=2\pi$. We require
the continuity of the wave functions, as well as a vanishing spin current
density at the junctions.\cite{Griffith,Xia,MPV,FMBP05a} The procedure is
similar to the case of a single outgoing lead which was described in detail in
Refs.~[\onlinecite{MPV},\onlinecite{FMBP05a}]. The results can be summarized
by the aid of two transmission matrices which acting on the incoming spinor
valued input wave functions provide the output:
\begin{eqnarray}
T^{\left( n\right) }
\begin{pmatrix}
f_{\uparrow} \\
f_{\downarrow}%
\end{pmatrix}
&=&%
\begin{pmatrix}
t_{\uparrow}^{n} \\
t_{\downarrow}^{n}%
\end{pmatrix}%
,
\end{eqnarray}%
with $n=1, 2.$ When the incoming electron is not perfectly spin-polarized, its
state should be described by a $2\times2$ density matrix $\rho_{in},$ we can
write:
\begin{equation}
\rho^{n}=T^{\left( n\right) }\rho_{in} \left(T^{\left(
n\right)}\right)^{\dagger},
\end{equation}
where $\rho^{1}$ and $\rho^{2}$ are the output density matrices in the
respective leads. The matrix elements of $T^{\left( 1\right)}$ and $T^{\left(
2\right)}$ can be calculated analytically for arbitrary geometry, but we found
that the spin polarizing properties of this device are most clearly seen for
the case when the outgoing leads are in a symmetric position, i.e., $\gamma
_{1}=2\pi -\gamma _{2}$. Here we will limit ourselves to this symmetric
geometry, yielding
\begin{eqnarray*}
T_{\uparrow \uparrow}^{\left( 1\right)
}\!\!&=&\!\frac{8qak}{y}e^{i\frac{\gamma _{2}}{2}}\left[ \cos
^{2}\frac{\theta} {2}\left( h_{1}+h_{2}\right) +\sin ^{2}\frac{\theta
}{2}\left( h_{1}^{\ast}
-h_{2}^{\ast }\right) \right],\notag \\
T_{\uparrow \downarrow}^{\left( 1\right)
}\!\!&=&\!\frac{8qak}{y}e^{i\frac{\gamma _{2}}{2}}\sin \frac{ \theta }{2}\cos
\frac{\theta }{2}\left[ \left( h_{1}+h_{2}\right) -\left(
h_{1}^{\ast }-h_{2}^{\ast }\right) \right] , \notag\\
T_{\downarrow \downarrow}^{\left( 1\right)
}\!\!&=&\!\frac{8qak}{y}e^{-i\frac{\gamma _{2}}{2}}\left[
\sin ^{2}\frac{\theta }{2}\left( h_{1}+h_{2}\right) +\cos ^{2}\frac{\theta }{%
2}\left( h_{1}^{\ast }-h_{2}^{\ast }\right) \right] \notag ,\\
T_{\downarrow \uparrow}^{\left( 1\right) }\!\!&=&\!e^{-i\gamma
_{2}}T_{\downarrow \uparrow}^{\left( 1\right) },
\end{eqnarray*}%
where
\begin{eqnarray}
h_{1} &=&-ake^{-i\frac{w}{2}\gamma _{2}}e^{iw\pi }\sin \left( q\left( 2\pi
-\gamma _{2}\right) \right) \sin \left( 2q\left( \pi -\gamma _{2}\right)
\right) ,  \notag \\
h_{2} &=&iqe^{-i\frac{w}{2}\gamma _{2}}\left[ e^{iw\pi }\sin \left( q\gamma
_{2}\right) -\sin \left( q\left( 2\pi -\gamma _{2}\right) \right) \right],\notag \\
y&=& ia^{3}k^{3}\left[ \sin \left( 2q\left( 3\pi -2\gamma _{2}\right)
\right) -2\sin \left( 2q\left( \pi -\gamma _{2}\right) \right) \right.  \notag \\
&&\left. -\sin \left( 2q\pi \right) \right] -2qa^{2}k^{2}\left[ \cos \left(
2q\left( 3\pi -2\gamma _{2}\right) \right) \right. \label{h1h2} \\
&&\left. +2\cos \left( 2q\left( \pi -\gamma _{2}\right) \right) \right]
+6qa^{2}k^{2}\cos \left( 2q\pi \right)  \notag \\
&&-12iq^{2}ak\sin \left( 2q\pi \right) +8q^{3}\left[ \cos \left( w\pi \right)
+\cos \left( 2q\pi \right) \right]. \notag
\end{eqnarray}%
Similarly, for the second output we obtain  $T_{\uparrow \uparrow}^{\left(
2\right)}=T_{\downarrow \downarrow}^{\left( 1\right)},$ $T_{\downarrow
\downarrow}^{\left( 2\right)}=T_{\uparrow \uparrow}^{\left( 1\right)},$
$T_{\uparrow \downarrow}^{\left( 2\right)}=-T_{\downarrow \uparrow}^{\left(
1\right)}$ and $T_{\downarrow \uparrow}^{\left( 2\right)}=-T_{\uparrow
\downarrow}^{\left( 1\right)}.$ This symmetry is related to the chosen
geometry $\gamma _{1}=2\pi -\gamma _{2}.$ We note that the reflection matrix
can also be calculated using the method described above, it turns out to be
diagonal in the $\{\left\vert \uparrow\right\rangle$, $\left\vert
\downarrow\right\rangle \}$ basis. We concentrate here on the transmission
properties of the ring and consider reflection as a loss in the efficiency of
spin transformation.

\begin{figure}[htb]
\begin{center}
\includegraphics*[width=8.2cm]{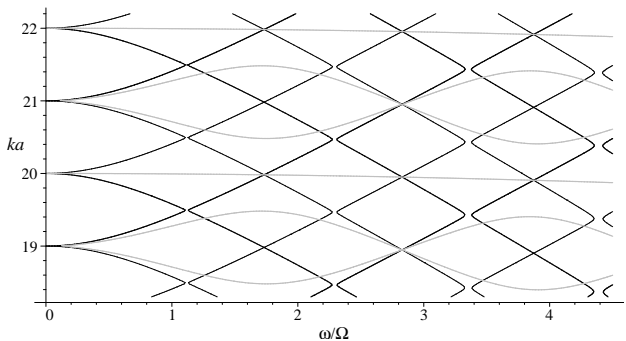}
\vspace{-0.5cm}
\end{center}
\caption{Determination of the parameter values corresponding to perfect
polarization: Eq.~(\ref{cos}) and Eq.~(\ref{sin}) with the plus (minus) sign
are satisfied along the gray and the thin (thick) black lines, respectively.
At each intersection of a black and a gray line the ring acts as a
perfect polarizing device. This figure corresponds to geometry given by
$\gamma_{1}=\pi/2,\gamma_{2}=3\pi/2.$} \label{hhfig}
\end{figure}

The most surprising physical consequence of our three terminal ring is its
ability to deliver polarized output beams of electrons. Considering a
completely unpolarized input, i.e., $\rho_{in}$ being proportional to the
identity, the outputs will be generally partially polarized that could be
detected by Faraday rotation experiments.\cite{KMGA05} However, we found that
properly chosen parameters lead to output polarizations as high as $100 \%$.
The relevant output density operators in this case should be projectors (apart
from the possible reflective losses):
\begin{equation}
\frac{1}{2}T^{\left(n \right)}\left(T^{\left( n \right)}\right)^{\dagger}
=\eta_{n}|\phi^{n}\rangle \langle \phi^{n}|. \label{projection}
\end{equation}
The  non-negative numbers $\eta_{1}$ and $\eta_{2}$ measure the efficiency of
the polarizing device, i.e., $\eta_{1}+\eta_{2}=1$ means a reflectionless
process. Direct calculation shows that, provided Eq.~(\ref{projection}) is
satisfied, the norms of the two outputs are equal, $\eta_{1}=\eta_{2}\equiv
\eta/2.$ Eq.~(\ref{projection}) is equivalent to requiring the determinants of
$T^{\left( n\right)} \left(T^{\left(n\right)}\right)^{\dagger}$ to vanish. We
found that these determinants are equal, and zero if $h_{1}\pm h_{2}=0$. Using
Eqs.~(\ref{h1h2}), these conditions can be formulated as
\begin{subequations}
\begin{eqnarray}
\cos \left( w\pi \right)  &=&\frac{\sin \left( q\gamma _{2}\right) }{\sin
\left( q\left( 2\pi -\gamma _{2}\right) \right) } \label{cos}, \\
\sin \left( w\pi \right)  &=&\mp \frac{ak}{q}\sin \left( 2q\left( \pi -\gamma
_{2}\right) \right),  \label{sin}
\end{eqnarray}
\end{subequations}
each of them lead to a $k-\omega$ relation as depicted in Fig.~\ref{hhfig} for
a representative example corresponding to $\gamma_{2}=3\pi/2$. The crossing
points of the gray (solution of Eq.~(\ref{cos})) and black (solution of
Eq.~(\ref{sin})) curves in Fig.~\ref{hhfig} are the parameters that can be
used in an experimental realization of our proposal to achieve perfectly
polarized outputs.
Similar figures can be drawn for arbitrary (symmetric) geometry. This implies
that there are lines in three dimensional $\{\gamma_{2},\omega/\Omega,ka\}$
space along which the ring polarizes a completely unpolarized input.

\begin{figure}[bh]
\begin{center}
\includegraphics*[width=8.2cm]{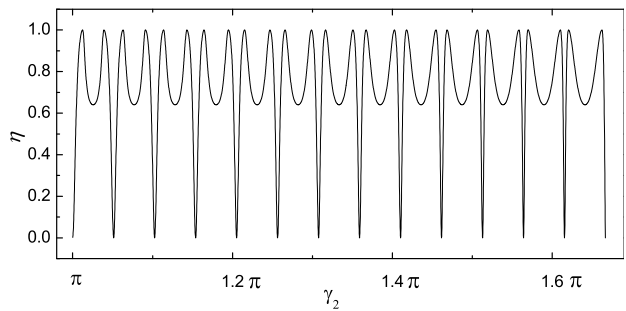}
\vspace{-0.4cm}
\end{center}
\caption{The transmission probability of a perfectly polarizing ring as a
function of $\gamma_{2}=2\pi-\gamma_{1}$. The parameter $ka$ changes in the
range of $[19.0,21.0],$ while $0<\omega/\Omega<5,$ and the plot corresponds to
the condition $h_{1}+h_{2}=0.$}\label{transfig}
\end{figure}

Now we can ask what the transmission probabilities  are, \emph{provided}
perfect polarization occurs. Fig.~\ref{transfig} shows that along a line
defined by $h_{1}+ h_{2}=0$, $\eta$ is a quasiperiodic function of
$\gamma_{2}$. A similar figure can be drawn for the condition $h_{1} -
h_{2}=0.$ As we can see, there are certain points (that is, parameter
combinations), where the transmission probability is unity. This shows that it
is possible to obtain $100\%$ spin polarized outputs from a perfectly
unpolarized input, even without reflective losses.

Now we turn to the investigation of the outgoing spinors which arise as a
consequence of the polarizing property of the ring. Clearly, these are the
eigenstates $|\phi^{n}\rangle $ of the transmitted density matrices
corresponding to the nonzero eigenvalues which are given by $ \eta _{1} =
\eta_{2}=128q^{2}a^{2}k^{2}{\left| h_{1} \right|^2}/{ \left\vert y\right\vert
^{2}}.$ Note that the quasiperiodic behavior of the transmission probability
$\eta= \eta _{1} + \eta _{2}$ seen in Fig.~{\ref{transfig}} is related to the
sine and cosine functions in $h_1$ and $y$. Focusing on the case of
$h_{1}+h_{2}=0$, the eigenstates of the respective transmitted density
matrices corresponding to the nonzero eigenvalues $\eta _{1}$ and $\eta_{2}$
read
\begin{equation}
|\phi^{1}\rangle_{+} = \left(
\begin{array}{c}
\sin \frac{\theta }{2} \\
-e^{-i\gamma _{2}}\cos \frac{\theta }{2}%
\end{array}
\right), \ \ |\phi^{2}\rangle_{+}=\left(
\begin{array}{c}
e^{-i\gamma _{2}}\cos \frac{\theta }{2} \\
\sin \frac{\theta }{2}%
\end{array}%
\right). \label{TTeigen}
\end{equation}
These results describe the connection between the strength of the spin-orbit
coupling (encoded in $\theta$), the geometry of the device and its polarizing
directions. We stress that this pair of spinors exhibits nontrivial
spatial-spin correlation being a signature of quantum
non-contextuality.\cite{HL03} However, note that they are in general not
orthogonal, their overlap is given by ${}_{+}\langle
\phi^{2}|\phi^{1}\rangle_{+}=i \sin \theta \sin \gamma_2.$ On the
distinguishability of nonorthogonal states, see Ref.~[\onlinecite{disting}].
Similarly, for $h_{1}-h_{2}=0$, we have:
\begin{equation}
|\phi^{1}\rangle_{-}=\left(
\begin{array}{c}
e^{i\gamma _{2}}\cos \frac{\theta }{2} \\
\sin \frac{\theta }{2}%
\end{array}
\right) ,\ \ |\phi^{2}\rangle_{-} = \left(
\begin{array}{c}
\sin \frac{\theta }{2} \\
-e^{i\gamma _{2}}\cos \frac{\theta }{2}%
\end{array}
\right). \label{TTeigen2}
\end{equation}

Considering the transmission matrices themselves, it is clear that under the
conditions given by Eqs.~(\ref{cos}~-b), their determinants also vanish. That
is, each $T^{(n)}$ has a zero eigenvalue, but -- due to the nonhermiticity --
its eigenspinors are not orthogonal. It can be verified that the eigenstates
corresponding to the nonzero eigenvalue coincide with $|\phi^{n}\rangle_{+}$
and $|\phi^{n}\rangle_{-}$, while the spinors annulled by the transmission
matrices $T^{(n)}|\phi^{n}_{0}\rangle=0$ have the following components:
\begin{equation}
|\phi^{1}_{0}\rangle_{+} = \left(
\begin{array}{c}
\cos \frac{\theta }{2} \\
\sin \frac{\theta }{2}
\end{array}
\right), \ \ |\phi^{2}_{0}\rangle_{+}=\left(
\begin{array}{c}
-\sin \frac{\theta }{2} \\
\cos \frac{\theta }{2}
\end{array}
\right) \label{Teigen}
\end{equation}
if $h_{1}+h_{2}=0$, and $|\phi^{1}_{0}\rangle_{-}=|\phi^{2}_{0}\rangle_{+},$
$|\phi^{2}_{0}\rangle_{-}=|\phi^{1}_{0}\rangle_{+}.$

This shows that if the conditions given by Eqs.~(\ref{cos}~-b) are
satisfied, the device acts similar to a Stern-Gerlach apparatus in the
sense that: 1) for unpolarized input, we have two different spin directions
(\ref{TTeigen}) in the outputs, 2) if we consider one of the eigenstates
(\ref{TTeigen}) as the input, its spin direction will not change in the appropriate
output, and 3) there are spinors given by Eq.~(\ref{Teigen}), for which the
transmission probability into a given output lead is zero. However, the
analogy is not perfect, the polarized spinors (\ref{TTeigen}) are not
orthogonal and the spinor which has zero probability to be transmitted through
a given lead is not equal to the eigenstate corresponding to the nonzero
eigenvalue of the other lead: $|\phi^{n}\rangle \neq
|\phi^{n^{\prime}}_{0}\rangle $ for $n \neq n^{\prime}.$ From this point of
view, an optical polarizing beam splitter \cite{L00,AAKJ04} with nonorthogonal
polarizing directions can be the closest analogue.

The present calculation was done for an idealized model system; in fact, our
intention was showing that the discussed polarizing effect -- in contrast to
previous proposals -- can be described in terms of pure Quantum Mechanics
(i.e., spin precession and interference), thus it is of importance from a
fundamental point of view, as well. On the other hand, there are results
showing that the approximations of our model (transport is ballistic and one
dimensional, i.e., the finite width of the ring-wire was not taken into
account) can give valid description of actual physical systems under specific
experimental situations. Currently, high mobility samples have become
available such that at cryogenic temperatures transport is found to be
ballistic over tens of microns. Similarly, phase coherence and spin coherence
lengths \cite{KA98} have been found up to 100 $\mu m$. Our narrow ring implies
the assumption of single mode propagation. Recently, it was found that the
finite width of the rings has a small effect on the loss of coherence of the
spin state; it has also been shown that in a multi-channel system the
modulation of the transmitted spin states survive and under specific
conditions the individual eigenchannel transmissions are very similar to the
ones found in single channel rings.\cite{FR04,SN04} A possible non-ideal
coupling to the leads can be described through effective tunnel barriers. But
in most of the current experimental systems the leads are connected in a
rather adiabatic way which makes the coupling very close to ideal.
\bigskip

In conclusion, we showed that a quantum ring with one input and two output
leads in the presence of Rashba-type SOI has remarkable similarities with a
Stern-Gerlach apparatus. Parameter values, within the experimentally feasible
range \cite{NATE97,NMT99,SK01} were identified when the three terminal ring
delivers perfectly polarized output beams of electrons without reflective
losses. We found that appropriate spin polarized input states are transmitted
without modification, but it is also possible to prepare inputs, for which the
transmission into a given lead is forbidden. Thus our paper describes a
realistic model in which spin sensitive quantum interference gives rise to
fundamental polarization effects as well as to nontrivial spatial-spin
correlations.

We note that similar rings can act as spintronic quantum gates \cite{FMBP05a}
or in the presence of an external magnetic field can be used also for spin
filtering.\cite{MPV} This points to the possibility to integrate spintronic
beam splitters, gates and filters that can serve as elementary building blocks
of a quantum network based on spin sensitive
devices.\cite{EBL03,SB03,YPS02,FHR01}

\bigskip
Acknowledgments: This work was supported by the Flemish-Hungarian Bilateral
Programme, the Flemish Science Foundation (FWO-Vl), the Belgian Science Policy
and the Hungarian Scientific Research Fund (OTKA) under Contracts Nos.~T48888,
D46043, M36803, M045596.

\end{document}